\documentstyle [twocolumn,epsf]{mn}
\newcommand{\mincir}{\raise
-3.truept\hbox{\rlap{\hbox{$\sim$}}\raise4.truept\hbox{$<$}\ }}
\newcommand{\magcir}{\raise
-3.truept\hbox{\rlap{\hbox{$\sim$}}\raise4.truept\hbox{$>$}\ }}
\newcommand{\minmag}{\raise
-3.truept\hbox{\rlap{\hbox{$<$}}\raise5.truept\hbox{$<$}\ }}
\newcommand{\be}{\begin{equation}}
\newcommand{\ee}{\end{equation}}
 \newcommand{\ba}{\begin{eqnarray}}
\newcommand{\ea}{\end{eqnarray}}
\newcommand{\brr}{\begin{array}}
\newcommand{\err}{\end{array}}
\newcommand{\bc}{\begin{center}}
\newcommand{\ec}{\end{center}}

\renewcommand{\baselinestretch}{1.0}

\title[SDSS Cluster Correlation Function]
{Modelling the two point correlation function of galaxy clusters in
the Sloan Digital Sky Survey}
\author[Basilakos \& Plionis ]{Spyros Basilakos$^{1}$ \& Manolis 
Plionis$^{1,2}$. \\
\vspace{0.1cm}
$^1$ Institute of Astronomy \& Astrophysics, National Observatory of Athens, 
I. Metaxa \& V. Pavlou, Palaia Penteli, 15236 Athens, Greece \\
$^2$ Instituto Nacional de Astrofisica, Optica y Electronica (INAOE)
Apartado Postal 51 y 216, 72000, Puebla, Pue., Mexico
}

\begin{document}

\maketitle

\begin{abstract}
We study the clustering properties of the recently compiled
SDSS cluster catalog using the two point 
correlation function in redshift space.
We divide the total SDSS sample into two richness subsamples, roughly 
corresponding to Abell $R\ge 0$ and APM clusters, respectively.
If the two point correlations are modeled as a 
power law, $\xi(r)=(r_{\circ}/r)^{\gamma}$, then 
the best-fitting parameters for the two subsamples are 
$r_{\circ}=20.7^{+4.0}_{-3.8} \; h^{-1}$ Mpc with 
$\gamma=1.6^{+0.4}_{-0.4}$ and $r_{\circ}=9.7^{+1.2}_{-1.2}$ with 
$\gamma=2.0^{+0.7}_{-0.5}\; h^{-1}$ Mpc, respectively.
Our results are consistent with the 
dependence of cluster richness to the cluster correlation length. 

Finally, comparing the SDSS cluster correlation function 
with that predictions from three flat cosmological models 
($\Omega_{\rm m}=0.3$) with dark energy (quintessence), we estimate
the cluster redshift space distortion 
parameter $\beta \simeq \Omega_{\rm m}^{0.6}/b_{\circ}$ and the cluster 
bias at the present time. 
For the $\Lambda$CDM case we find $\beta=0.2^{+0.029}_{-0.016}$, which is
in agreement with the results based on the large scale cluster
motions.

{\bf Keywords:} galaxies: clusters: general - cosmology: theory - 
large-scale structure of universe 
\end{abstract}

\vspace{1.0cm}

\section{Introduction}

Galaxy clusters 
occupy a special position in the hierarchy of cosmic structure
formation, being the largest gravitationally collapsed objects 
in the universe. Therefore, they appear to be ideal tools for 
testing theories of structure formation as well as studying 
large-scale structure.
The traditional indicator of clustering, the 
cluster two-point correlation function, is a fundamental statistical test
for the study of the cluster distribution and is relatively
straightforward to measure from observational data. 

Indeed, many authors based on optical and X-ray data have shown that 
the large scale clustering pattern of galaxy clusters is well
described by a power law, $\xi(r)=(r_{\circ}/r)^{\gamma}$, with 
$\gamma=1.6-2$. The correlation length $r_{\circ}$ lies in the interval
$r_{\circ}=13-25h^{-1}$Mpc, depending on the cluster richness as 
well as the analyzed sample (cf. Bahcall \& Soneira 1983; Klypin \&
Kopylov 1983; Lahav et al. 1989; 
Bahcall \& West 1992; Peacock \& West 1992; 
Dalton et al. 1994; Nichol, Briel \& Henry 1994; 
Croft et al. 1997; Abadi, Lambas \&
Muriel 1998; Borgani, Plionis \& Kolokotronis 1999; Collins et al. 2000; 
Tago et al. 2002; Moscardini, Matarrese \& Mo 2001; Gonzalez, 
Zaritsky \& Wechsler 2002). 
However, a serious issue here is how the galaxy clusters trace the
underlying mass distribution. The cluster distribution traces scales that 
have not yet undergone the non-linear
phase of gravitationally clustering and thus simplifying their connection 
to the initial conditions of cosmic structure formation. 
Galaxy clusters is strong biased with respect to the matter distribution
(e.g. Peacock \& Dodds 1994 and references therein).

In this paper we utilize the recently completed SDSS CE 
cluster catalog (Goto et al. 2002) in order:
(i) to study the two point correlation function 
in redshift space and
(ii) to calculate the relative cluster bias at the present time comparing the 
observational results with those derived from three flat cosmological 
models with dark energy (quintessence). The structure of the 
paper is as follows. The observed dataset and its 
measured correlation function are presented in section
2. In section 3 we give a brief account 
of the method used to estimate the predicted 
correlation function in different CDM spatially 
flat cosmologies. The linear growth rate of clustering in quintessence
cosmological models can be found in section 4, while in section 5 we
fit the SDSS cluster clustering to different cosmological and biasing models.
Finally, we draw our conclusions in section 6.

\section{Estimation of the SDSS cluster correlation function}
\subsection{Cluster catalogue}
In this work we use the recent SDSS CE cluster catalog 
(Goto et al. 2002), which contains 2770 and 1868 galaxy clusters 
in the North ($145.1^{\circ} <\alpha< 236.0^{\circ}$, 
$-1.25^{\circ} <\delta< 1.25^{\circ}$) and South 
($350.5^{\circ} <\alpha< 56.61^{\circ}$, 
$-1.25^{\circ} <\delta< 1.25^{\circ}$) 
slices respectively, 
covering an area of $\sim 400$deg$^2$ in the sky.
Redshifts are converted to proper distances 
using a spatially flat cosmology with 
$H_{\circ}=100h\,$km$\,$s$^{-1}\,$Mpc$^{-1}$ and 
$\Omega_{\rm m}=0.3$. 
The cluster redshifts are estimated
using the color information by identifying the bin in $g-r$ which 
has the largest number of galaxies around the color prediction of
elliptical galaxies (Fukugita, Shimasaku, Ichikawa 1995)
at different redshifts (which define the different
$g-r$ bins). Due to the fact that the true and estimated
redshifts are better correlated for $z<0.3$
(Goto et al. 2002), we will limit our analysis within this redshift range,
corresponding to a limiting distance of $r_{\rm max}\le 836h^{-1}\,$Mpc.

In Fig. 1, we present the estimated (histogram) 
and the expected for a volume limited sample (solid line), number 
of the SDSS clusters as a function of redshift. It is evident that the
number of SDSS clusters appears to follow the equal-volume 
$\propto r^3$ law out to
$z\sim 0.23$, a fact corroborated also from the standard
Kolmogorov-Smirnov (KS) test which gives probability of 
consistency between model and observations 
(up to $z \le 0.23$) of ${\cal P}_{\rm KS}\simeq0.43$.
Therefore, this SDSS cluster sample is the only to-date sample that is
volume limited to such a large distance and can thus play an important
role in large scale structure studies.

\begin{figure}
\mbox{\epsfxsize=9.4cm \epsffile{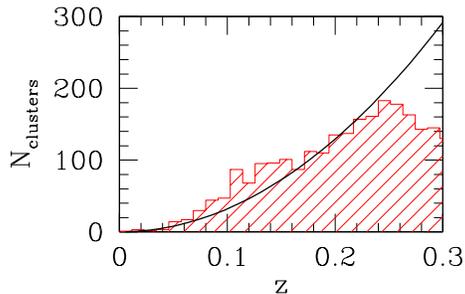}}
\caption{The estimated (histogram) and the expected (line) number of the
SDSS clusters as a function of redshift.} 
\end{figure}

We apply the cluster correlation function analysis using  
clusters of two richness class: (a) $N_{\rm gal}\ge 30$ members
(roughly corresponding to Abell $R\ge 0$; hereafter $S_{1}$ sample) and
(b) $N_{\rm gal} \ge 20$ members (roughly corresponding to APM clusters; 
hereafter $S_{2}$ sample). These two subsamples contains 200 and 524
entries with corresponding mean densities of 
$n_{\rm S_{1}} (\le z_{\rm max})\simeq 8.42(\pm 0.06) 
\times 10^{-6}h^{3}\,$Mpc$^{-3}$ 
and 
$n_{\rm S_{2}} (\le z_{\rm max})\simeq 2.20(\pm 0.10) 
\times 10^{-5}h^{3}\,$Mpc$^{-3}$, 
giving rise to inter-cluster separations of the order of
$d_{\rm S_{1}}\sim 49.15 \pm 2.56$ $h^{-1}$Mpc and
$d_{\rm S_{2}}\sim 35.66 \pm 2.18$ $h^{-1}$Mpc, respectively.

\subsection{SDSS cluster correlations}
We estimate the redshift space correlation function 
using the estimator described by Hamilton (1993):
\be
\xi_{\rm S_{j}}(r)=4\frac{N_{DD} \langle N_{RR} 
\rangle}{\langle N_{DR}\rangle^{2}}-1
\ee 
where $j=1, 2$ and $N_{DD}$ is the number of 
cluster pairs in the interval $[r-\Delta r,r+\Delta r]$. While,
$\langle N_{RR} \rangle$ and $\langle N_{DR} \rangle$ is the average, over 
10000 random simulations with the same properties as the real data 
(boundaries and redshift selection function),
cluster-random and random-random pairs, respectively. The random
catalogues were constructed by randomly reshuffling the angular
coordinates of the clusters (within the limits of the catalogue), 
while keeping the same redshifts and thus exactly the same redshift
selection function as the real data. 

Note that in order to take into account the possible systematic 
effects (eg. fraction of high-$z$ clusters missed by the
finding algorithm due to SDSS magnitude limit) in the 
different cluster subsamples 
we generate random catalogs, utilized the individual distance 
distribution of each subsample and not the overall SDSS cluster 
selection function. We compute the errors on $\xi_{\rm S_{j}}(r)$
from 100 bootstrap re-samplings of the data (Mo, Jing 
\& B$\ddot {\rm o}$rner 1992).   

\begin{figure}
\mbox{\epsfxsize=8cm \epsffile{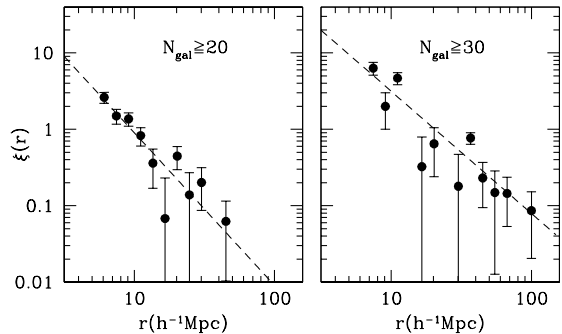}}
\caption{The spatial two-point correlation function (points) in redshift space
for the $\rm S_{1}$ (Abell R=0 richness-right panel) and 
$\rm S_{2}$ (APM richness-left panel) samples. 
The error bars are estimated using the bootstrap procedure. The dashed lines
represent the best-fitting power low
$\xi_{\rm S_{j}}(r)=(r_{\circ}/r)^{\gamma}$ (see parameters in table
1).} 
\end{figure}

\begin{table*}
\caption[]{Results of the correlation function analysis 
for clusters of the two richness class (${\rm S}_{1}$ and $\rm S_{2}$ samples).
Errors of the fitted parameters represent $2\sigma$ uncertainties.
Finally, the $r_{\circ}$ has units of $h^{-1}\,$Mpc.}

\tabcolsep 9pt
\begin{tabular}{ccccc} 
\hline
Sample & No. of clusters& $r_{\circ}$ & $\gamma$ &$r_{\circ}(\gamma=1.8)$ \\ \hline \hline 
   $\rm S_{1}$  &  200 &  $20.7^{+4.0}_{-3.8}$& $\gamma=1.6^{+0.4}_{-0.4}$& 
$r_{\circ}=19.8^{+2.9}_{-3.2}$ \\
   $\rm S_{2}$  &  524 &  $9.7^{+1.2}_{-1.2}$& $\gamma=2.0^{+0.7}_{-0.5}$& 
$r_{\circ}=9.8^{+1.2}_{-1.3}$ \\
\end{tabular}
\end{table*}

We apply the correlation analysis to the $\rm S_{1}$ and $\rm S_{2}$ 
subsamples evaluating $\xi_{\rm S_{j}}(r)$ in logarithmic intervals.
In Fig. 2, we present the estimated two point redshift
correlation function 
(dots), divided according to richness class; strong clustering is evident.
The dashed lines correspond to the best-fitting power law model 
$\xi_{\rm S_{j}}(r)=(r_{\circ}/r)^{\gamma}$, which is determined 
by the standard $\chi^{2}$ minimization procedure in which each
correlation point is weighted by its error$^{-1}$. The fit has 
been performed taking into account
bins with $r\ge 5 \;h^{-1}$Mpc in order to avoid the signal from small, non-linear, 
scales while we have used no upper $r$ cut-off (due to our
$\sigma^{-1}$ weighting
scheme, our results remain robust by varying the upper $r$
limit within the 25 to 100 $h^{-1}$ Mpc range).

In Fig. 3 we present the iso-$\Delta \chi^{2}$ contours 
(where $\Delta \chi^{2}=\chi^{2}-\chi_{\rm min}^{2}$) in the $\gamma-r_{\circ}$
plane. The $\chi_{\rm min}^{2}$ is the absolute minimum value of 
the $\chi^{2}$. 
The contours correspond to $1\sigma$ ($\Delta \chi^{2}=2.30$)
and $2\sigma$ ($\Delta \chi^{2}=6.17$) uncertainties, respectively. In
the insert of Fig. 3 we show the variation of $\Delta \chi^{2}$ 
around the best fit, once we marginalize with respect to the other
parameter, while in Table 1, we list all the relevant information. 
For the $\rm S_{1}$ cluster subsample (Abell $R\ge 0$ richness)
the best fitted clustering parameters are 
$r_{\circ}=20.7^{+4.0}_{-3.8}h^{-1}$Mpc and $\gamma=1.6^{+0.4}_{-0.4}$ 
which are in very good agreement with the values 
$r_{\circ}=20.6\pm 1.5h^{-1}$Mpc and $1.5\pm 0.2$ derived 
by Peacock \& West (1992)\footnote{The robustness of our results to 
the fitting procedure was tested using different bins (spanning from
10 to 20) and we found very similar clustering results.}. 
Results for the $\rm S_{2}$ subsample 
(APM richness) $r_{\circ}=9.7^{+1.2}_{-1.2}h^{-1}$Mpc 
and $\gamma=2.0^{+0.7}_{-0.5}$
can be compared with those obtained by Dalton et al (1994);
Bahcall \& West (1992) and recently, from Plionis \& Basilakos (2002), 
based on the APM cluster catalog. They found a somewhat greater
correlation length $r_{\circ} \simeq 12 - 13 \; h^{-1}$ Mpc. 
We can further estimate an upper limit of the 
correlation length using the expression between the $r_{\circ}$ and
the mean cluster separation of Bachall \& Burgett (1986), as modified
by Bahcall \& West (1992):
$r_{\circ,{\rm S_{2}}} \simeq 0.4d_{\rm S_{2}}\simeq 14.2 \; h^{-1}$ Mpc 
(see also Dalton et al. 1994 and Croft et al 1997). 

\begin{figure}
\mbox{\epsfxsize=8cm \epsffile{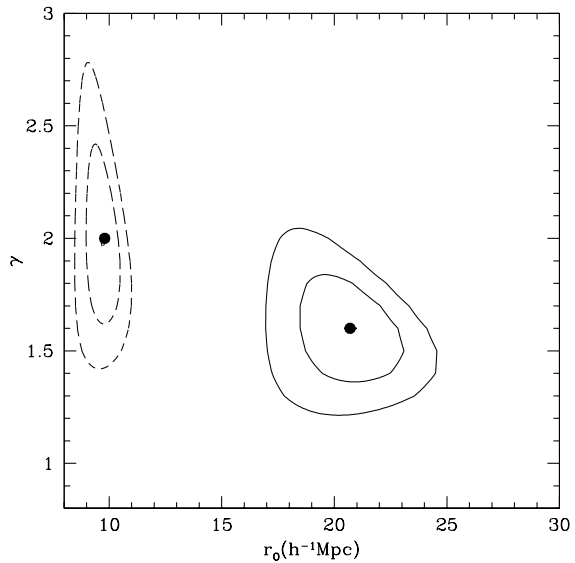}}
\caption{Iso-$\Delta \chi^{2}$ contours  in the $\gamma$-$r_{\circ}$ 
parameter space for the $\rm S_{1}$ 
(continuous line) and $\rm S_{2}$ (dashed line) samples.
In the insert we show the variation of $\Delta \chi^{2}$ around the best
fit once we marginalize with respect to the other parameter.}
\end{figure}

In order to directly compare the correlation lengths 
of the two subsamples, we fixed the correlation function
slope to its nominal value of $\gamma=1.8$ and we found
$r_{\circ}=19.8^{+2.9}_{-3.2}$ and  
$r_{\circ}=9.8^{+1.2}_{-1.3}$ respectively (see last column of Table 1.).
It is clear that the correlation length increases with cluster richness, 
as expected from the well-known richness dependence of 
the correlation strength.
\begin{figure}
\mbox{\epsfxsize=8cm \epsffile{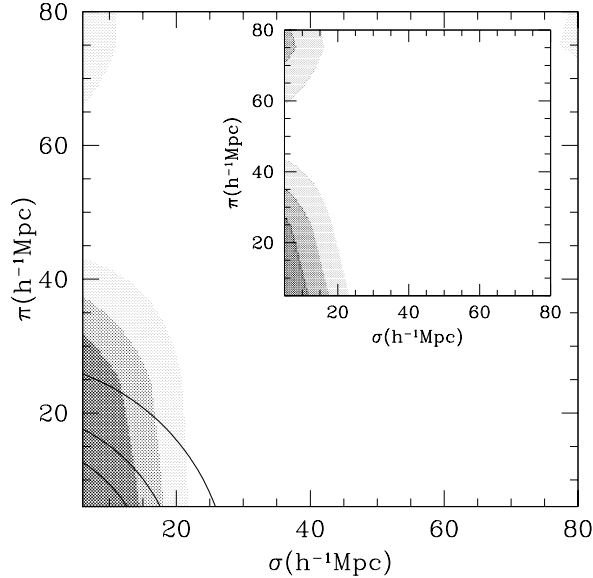}}
\caption{The radial and tangential anisotropy of the
two point correlation function of the $S_1$ SDSS cluster subsample
(while in the insert we show the results of the $S_2$ subsample).
The transitions between different shadings
correspond to fixed values of $\xi(\sigma,\pi)=1, 0.75, 0.5$.
The lines corresponds to the expected contours, using
$\xi(\sigma, \pi)=\left(r_0/(\sigma^2+\pi^2)\right)^\gamma$.
}  
\end{figure}

Finally, we have investigated the isotropy of the clustering
signal for both subsamples, by examining the radial and tangential
component of the SDSS correlation function
$\xi(\sigma,\pi)$, with $\pi$ the line-of-sight separation and
$\sigma$ the perpendicular 
component of the cluster separation $r$ (cf. Efstathiou et al. 1992).
We have used bins of 20 $h^{-1}$ Mpc width and in 
Fig. 4 we present the $\xi(\sigma, \pi)$ for
both subsamples.
It is evident that the $\xi(\sigma,\pi)$ contours are elongated 
along the line-of-sight direction, $\pi$, up to $\sim 40 \; h^{-1}$ Mpc.
However, we suspect that this is not an indication of systematic 
effects related to line-of-sight projections but rather 
due to the extremely small width of the survey area ($2.4$ degrees in
the declination direction which corresponds to $\sim 25 \; h^{-1}$ Mpc
at a redshift of $\sim 0.25$), a fact that gives
predominance to superclusters elongated along the line-of-sight with
respect to those in the perpendicular direction (see Jing,
Plionis \& Valdarnini 1992 for the effects of superclusters elongated
along the line-of-sight). Below we investiate our suspesion using the 
Hubble volume $\Lambda$CDM simulation (cf. Frenk et al 2000).

\subsection{Testing the robustness of the SDSS cluster correlations}
In order to test whether it is possible to recover the true underline cluster 
correlations from a survey with the geometrical characteristics,
selection function and richness of the SDSS, 
we have used the $\Lambda$CDM Hubble volume cluster catalogues (Colberg et
al 2000).
As an example, we present in Fig.5 the underline  
$S_1$-like cluster correlation function, estimated from the whole volume 
(continuous line)
and the mean of 6 mock $S_1$ SDSS cluster samples (which contain around 200
clusters each). 
The mean clustering length of the SDSS mock samples is $r_{\circ}
\simeq 18.5 \; h^{-1}$ Mpc while that of the underline cluster
population is $r_{\circ} \simeq 19.2 \; h^{-1}$ Mpc.
It is evident that the SDSS survey is adequate to
recover the underline clustering signal, albeit with a scatter of
$\sigma(\xi)/\xi \simeq 0.3$ at separations, for example of, $r\simeq 15
\;h^{-1}$ Mpc.

\begin{figure}
\mbox{\epsfxsize=8cm \epsffile{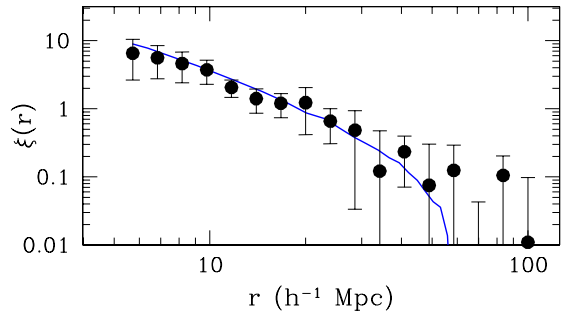}}
\caption{Comaprison of the underline $\Lambda$CDM $S_1$-like cluster
  correlation function (continuous line) with the mean of 6 SDSS mock samples.}
\end{figure}

Furthermore, we address the issue of the observed anisotropies
along the line of sight (see Fig. 4) by searching whether mock
obervers show similar $S_1$-like 
clustering elongations along either their $\pi$ or
$\sigma$ directions. We quantify this anisotropy by the
ratio, ${\cal R}$,  of the $\xi(\sigma,\pi)$ 
in the bins $(\pi, \sigma) = (0-20, 20-40) \; h^{-1}$ Mpc
(hereafter $\xi_{1,2}$) and $(\pi, \sigma) = (20-40,
0-20) \; h^{-1}$ Mpc (hereafter $\xi_{2,1}$).

We have selected 114 independent 
mock SDSS surveys (by spanning the $z$ coordinate
axis of the simulation) and we have found that in the majority
of the cases ($\sim$ 60\%) the value of ${\cal R}$
is larger than unity, indicating a predominance of 
anisotropies along the $\pi$ direction, while only $\sim$40\% 
of the cases it is less than one, indicating 
anisotropies along the $\sigma$ direction. 

\begin{figure}
\mbox{\epsfxsize=8cm \epsffile{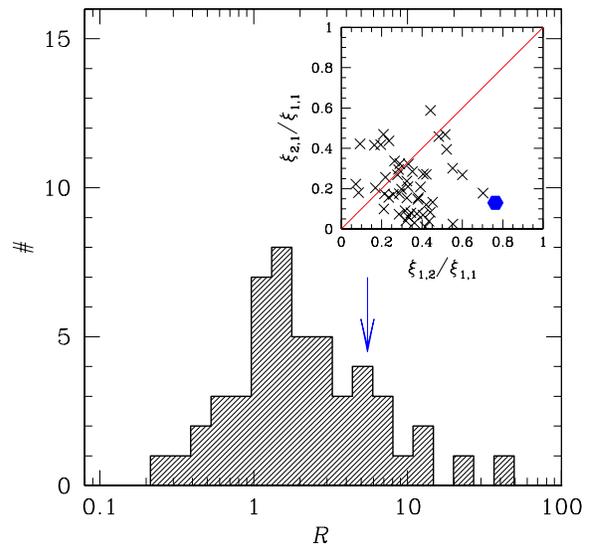}}
\caption{The ${\cal R}=\xi_{1,2}/\xi_{2,1}$ for 50 mock $S_1$ like
SDSS surveys for which $\xi_{1,2}$ or $\xi_{2,1}$ is larger than 0.4.
Note that ${\cal R}>1$ values indicate anisotropies along the $\pi$
direction, while the arrow shows the observed SDSS ${\cal R}$
value. The insert plot shows the amplitude of the
anisotropies, with respect to the central value ($\xi_{1,1}$).
Note that the continuous line divides
anisotropies along the $\pi$ direction (right of the line) and
anisotropies along the $\sigma$ direction (left of the line).}
\end{figure}
\begin{figure}
\mbox{\epsfxsize=8cm \epsffile{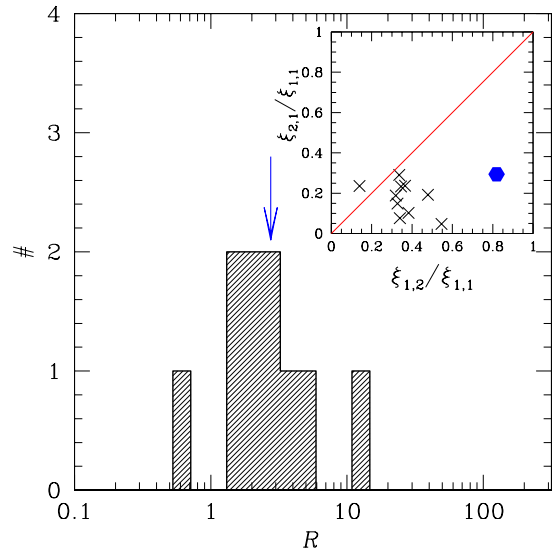}}
\caption{As in Fig.6 but for the $S_2$ sample.}
\end{figure}
We then investigated the amplitude of these $S_1$ clustering 
anisotropies by deriving the 
distribution of the ${\cal R}$ values for those observers that see
relatively high correlation values; $\xi_{1,2}>0.4$ and
$\xi_{2,1}>0.4$ (Fig. 6). In total we find 50 such observers out of
which 37 (74\%) show elongation along their line-of-sight.
Therefore, it is evident, also 
due to the tail towards large ${\cal R}$ values, 
that there are systematic anisotropies along the line-of-sight,
which of course could be only due to the geometric characteristics of
the mock cluster distribution.
Furthermore in the insert of figure 6 we plot a scatter diagram
between the normalized, by the value of $\xi(\sigma,\pi)$ at the
first bin (ie., $0<\sigma, \pi<20 \; h^{-1}$ Mpc), values of
$\xi_{1,2}$ and $\xi_{2,1}$. It is apparent that the 
observed SDSS value (filled point) is 
roughly consistent with of the simulation derived values, although it
appears to be an extremum. 
This indicates that the major part of the observed anisotropy is 
indeed due to the geometrical characteristics of the survey, as we
have anticipated in the previous subsection, however we cannot exclude that 
some small contribution from intrinsic systematics effects, 
like projection effects (cf. Sutherland 1988), could be present.

A similar analysis of the $S_2$ richness cluster correlations has
shown that the percentage of mock observers having significant clustering
anisotropies along the $\pi$ direction is significantly higher than
those seeing anisotropies along the $\sigma$ direction (9 out of 10), 
however the amplitude of these anisotropies appear to be 
lower than in the observed case. In Fig. 7 we show the corresponding
${\cal R}$ distribution and the scatter plot, in which it is evident
that the amplitude of the true SDSS clustering anisotropy in the $\pi$
direction is quite larger than what expected due to the survey
geometrical characteristics. 

Therefore, we conclude that in the case of the $S_1$ cluster
correlations there is no significant evidence for contamination by
projection effects, while in the case of the $S_2$ correlations we do
have such indications. However, in order
to perform an {\em a posteriori} correction of 
$\xi(r)$ (cf. Efstathiou 1992), for projection effects it would be
necessary to disentangle first the effects of the survey geometry, a
task which at the present time is out of the scope of this work.
Therefore, we caution the reader that all results based on the $S_2$
cluster sample could be affected by the above mentioned systematic
effect.

\section{Model cluster correlations}
It is well known (cf. Kaiser 1984; Benson et al. 2000) that assuming
linear biasing the mass-tracer and dark-matter correlations, at some redshift
$z$, are related by:
\be
\label{eq:spat}
\xi_{\rm model}(r,z)=\xi_{\rm DM}(r,z)b^{2}(z) \;\;, 
\ee
where $b(z)$ is the bias redshift evolution function. 
In the present work we have used the 
so called test particle bias model described by   
Nusser \& Davis (1994), Fry (1996) and Tegmark \& Peebles (1998). In 
this case the evolution of the correlation bias is developed  
assuming that only the test particle fluctuation field is 
related proportionally to that of the underling mass. 
Therefore, the bias factor as a function of redshift 
is
\be\label{eq:fry}
b(z)=1+\frac{(b_{\circ} - 1)}{D(z)} \;\; ,
\ee
with $b_{\circ}$ being the bias at the present time and
$D(z)$ the linear growth rate of clustering (described in section 4).
It has been found (Bagla 1998) that, in the interval $0\le z\le 1$, the above 
formula represents well the evolution of bias. Furthermore, the more
accurate linear bias evolution model given by Basilakos \& Plionis
(2001; 2003) is also very similar to the model of eq.(\ref{eq:fry})
within $z\le 1$.  

We quantify the evolution of clustering with epoch
presenting the spatial correlation function of the mass 
$\xi_{\rm DM}(r,z)$ as the Fourier transform of the 
spatial power spectrum $P(k)$:
\be
\xi_{\rm DM}(r,z)=
D^{2}(z)\frac{1}{2\pi^{2}}\int_{0}^{\infty} k^{2}P(k) 
\frac{{\rm sin}(kr)}{kr}{\rm d}k \;\;,
\ee
where $k$ is the comoving wavenumber. 

As for the power spectrum, we consider that of CDM models, 
where $P(k) \approx k^{n}T^{2}(k)$ with
scale-invariant ($n=1$) primeval inflationary fluctuations. 
We utilize the transfer function 
parameterization as in Bardeen et al. (1986), with the approximate
corrections given by Sugiyama's (1995) formula:

$$T(k)=\frac{{\rm ln}(1+2.34q)}{2.34q}[1+3.89q+(16.1q)^{2}+$$
$$(5.46q)^{3}+(6.71q)^{4}]^{-1/4} \;\; .$$
with
\be 
q=\frac{k}{\Omega_{\circ}h^{2}{\rm exp}[-\Omega_{b}-(2h)^{1/2}
\Omega_{b}/\Omega_{\circ}]}
\ee
where $k=2\pi/\lambda$ is the wavenumber in units of 
$h$ Mpc$^{-1}$ and $\Omega_{b}$ is the baryon density. 

In the present analysis we consider flat models
with cosmological parameters that fit the 
majority of observations, ie.,
$\Omega_{\rm m}+\Omega_{Q}=1$, $H_{\circ}=100h $km s$^{-1}$ 
Mpc$^{-1}$
with $h\simeq 0.7$ (cf. Freedman et al. 2001; Plionis 2002;
Peebles and Ratra 2002 and references therein), baryonic density 
parameter $\Omega_{\rm b} h^2 \simeq 0.02$ (e.g. 
Olive, Steigman \& Walker 2000; Kirkman et al 2003) 
and a CDM shape parameter $\Gamma=0.17$. 
In particular, we investigate 3 spatially 
flat low-$\Omega_{\rm m}=0.3$ cosmological models 
with negative pressure and values of $w=-1$ ($\Lambda$CDM),
$w=-2/3$ (QCDM1) and $w=-1/3$ (QCDM2).
Note that all the cosmological models 
are normalized to have fluctuation amplitude, in a sphere of
8 $h^{-1}$Mpc radius, of
$\sigma_{8}=0.50 (\pm0.1) \Omega_{\rm m}^{-\gamma}$ 
(Wang \& Steinhardt 1998) with 
$\gamma=0.21-0.22w+0.33\Omega_{\rm m}$. 

\section{The linear growth rate of clustering}
For homogeneous and isotropic flat cosmologies, driven by 
non relativistic matter and an
exotic fluid (quintessence models) with equation of state, $p_{Q}=w\rho_{Q}$
and $-1 \le w <0$, the Friedmann field equations can be 
written as:
\be\label{eq:11}
H^{2}=\left( \frac{\dot{\alpha}}{\alpha} \right)^{2}=
\frac{8\pi G}{3}(\rho_{\rm m}+\rho_{Q})
\ee
and 
\be
\frac{\ddot{\alpha}}{\alpha}=-4\pi G
[(w+\frac{1}{3})\rho_{Q}+\frac{1}{3}\rho_{\rm m}] \;\;,
\ee
where $\alpha(t)$ is the scale factor, 
$\rho_{\rm m} \propto \alpha(t)^{-3}$ is the matter density and 
$\rho_{Q} \propto \alpha(t)^{-3(1+w)}$ is the dark 
energy density. 

The time evolution equation for the mass density contrast, 
modeled as a pressureless fluid has general solution of the growing mode
(Peebles 1993):
\be\label{eq:11}
\ddot{D}+2H(t)\dot{D}=4\pi G \rho_{\rm m} D\;\; ,
\ee
where dots denote derivatives with respect to time.
From the equations describing the Friedmann model, it follows that
$\dot{H}+H^{2}=-4\pi G[(w+1/3)\rho_{Q}+(1/3)\rho_{\rm m}]$.
Differentiating this relation and using $\dot{\rho_{\rm m}}=-3H\rho_{\rm m}$
$\dot{\rho_{Q}}=-3(1+w)H\rho_{Q}$ we obtain
\be\label{eq:15}
\ddot{H}+2H\dot{H}=4\pi G(1+w)(w+\frac{1}{3})\rho_{Q} H
+4\pi G \rho_{\rm m} H\; .
\ee
Therefore, it turns out that if $w=-1$ ($\Lambda$CDM) or $w=-1/3$ (QCDM2)
then $H(t)$ is a decaying mode of eq.(\ref{eq:11}).
In that case, the growing solution (Peebles 1993) as a function 
of redshift is:
\be\label{eq:24}
D(z)=\frac{5\Omega_{\rm m} E(z)}{2}\int^{\infty}_{z} \frac{(1+x)}{E^{3}(x)} 
{\rm d}x\;\;. 
\ee
where we have used the following expressions:
\be\label{eq:4}
E(z)=\left[ \Omega_{\rm m}(1+z)^{3}+\Omega_{Q}(1+z)^{3(1+w)}\right]^{1/2}
\ee
\be\label{eq:5}
\frac{dt}{dz}=-\frac{1}{H_{\circ}E(z)(1+z)} \;\;.
\ee

The Hubble parameter is given by: $H(z)= H_{\circ} E(z)$,
while $\Omega_{\rm m}= 8\pi G \rho_{o}/3H_{o}^{2}$ 
(density parameter), $\Omega_{Q}= 8\pi G \rho_{Q}/3H_{o}^{2}$ 
(dark energy parameter) at the present time, which satisfy
$\Omega_{\rm m}+\Omega_{Q}=1$
and finally $H_{\circ}$ is the Hubble constant. 
In addition to $\Omega_{m}(z)$ also $\Omega_{Q}(z)$ could evolve 
with redshift as
\be
\Omega_{\rm m}(z)=\frac{\Omega_{\rm m} (1+z)^{3}}{E^{2}(z)} 
\ee
and
\be 
\Omega_{Q}(z)=\frac{\Omega_{Q} (1+z)^{3(1+w)}}{E^{2}(z)} \;\; .
\ee

It is interesting to mention that in a 
flat low-$\Omega_{\rm m}$ with $w=-1/3$ model, the equation of 
state $p_{Q}=-(1/3)\rho_{Q}$ 
leads to the same growing mode as in an open universe, despite 
the fact that this quintessence model has a spatially flat geometry!
Therefore, as the time evolves with redshift, 
utilizing equations (\ref{eq:5}), 
(\ref{eq:4}) and the relation
\be
4\pi G \rho_{\rm m}=\frac{3H_{\circ}^{2}}{2} \Omega_{\rm m}(1+z)^{3}\;,
\ee
then the basic differential equation for the evolution of the linear
growing mode takes the following form:
\be\label{eq:gen2}
\frac{{\rm d}^{2} D}{{\rm d} z^{2}}+P(z)\frac{{\rm d} D}{{\rm d} z}+
Q(z)D=0 \;\; 
\ee
with basic factors, 
\be\label{eq:ff1}
P(z)=-\frac{1}{1+z} +\frac{1}{E(z)}\frac{{\rm d}E(z)}{{\rm d}z}
\ee
and
\be
Q(z)=\frac{3\Omega_{\rm m} (1+z)}{2E^{2}(z)} \;\; .
\ee
We find that eq.(\ref{eq:gen2}) has a decaying 
solution of the form $D_{1}(z)=(1+z)^{n}$ 
only for $w=-2/3$, with $n=3/2$. The second independent 
solution of eq.(\ref{eq:gen2}) can be found easily from the following expression:
\be 
D(z)=D_{1}(z)\int_{z}^{\infty} \frac{ {\rm d} x } { D_{1}^{2}(x) } \;{\rm exp}
\left[ -\int^{x} P(t) {\rm d} t \right] 
\ee
which finally leads to the following growing mode:
\be\label{eq:ff2}
D(z)=(1+z)^{3/2}\int_{z}^{\infty} \frac{{\rm d}x}{(1+x)^{2}E(x)} \;\;.
\ee  

\begin{figure}
\mbox{\epsfxsize=8cm \epsffile{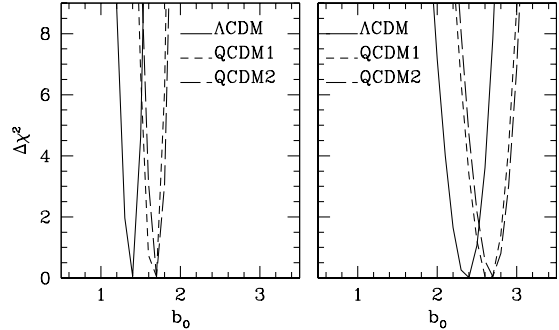}}
\caption{The variance $\Delta \chi^{2}$ around the best fit
$b_{\circ}$ value for various cosmological models. The left and the 
right panel correspond to the $\rm S_{2}$ and $\rm S_{1}$ samples, 
respectively.} 
\end{figure}

\begin{table*}
\caption[]{List of the theoretical clustering model fitting analysis. 
Note that the errors of the fitted parameters represent $2\sigma$ 
uncertainties.}

\tabcolsep 9pt
\begin{tabular}{cccc} 
\hline Index & $b_{\circ}$& $\beta$& $K(\beta)$ \\ \hline \hline 
$\Lambda$CDM-$\rm S_{1}$ &  $2.4^{+0.2}_{-0.3}$&  $0.20^{+0.029}_{-0.016}$& $1.14^{+0.02}_{-0.01}$\\
 QCDM1-$\rm S_{1}$  &  $2.6^{+0.3}_{-0.2}$&  $0.19^{+0.016}_{-0.020}$& $1.13^{+0.01}_{-0.01}$ \\
 QCDM2-$\rm S_{1}$ &  $2.7^{+0.2}_{-0.3}$&  $0.18^{+0.023}_{-0.013}$ &
 $1.12^{+0.02}_{-0.01}$ \\ \\

$\Lambda$CDM-$\rm S_{2}$  &  $1.4^{+0.1}_{-0.1}$&  $0.35^{+0.027}_{-0.023}$& $1.26^{+0.02}_{-0.02}$ \\
 QCDM1-$\rm S_{2}$  &  $1.7^{+0.1}_{-0.1}$&  $0.29^{+0.019}_{-0.019}$& $1.21^{+0.01}_{-0.01}$ \\
 QCDM2-$\rm S_{2}$  &  $1.7^{+0.1}_{-0.1}$&  $0.30^{+0.019}_{-0.016}$& $1.22^{+0.01}_{-0.01}$ \\
\end{tabular}
\end{table*}

\section{The SDSS cluster biasing}

In order to quantify the cluster bias at the present time
we perform a standard $\chi^{2}$ 
minimization procedure (described before) between the measured  
correlation function of the SDSS galaxy clusters 
with those expected in our spatially flat cosmological models
\be
\chi^{2}(b_{\circ})=\sum_{i=1}^{n} \left[ \frac{\xi_{\rm S_{j}}^{i}(r)-
\xi_{\rm model}^{i}(r,b_{\circ})}
{\sigma^{i}}\right]^{2} \;\;,
\ee 
where $\sigma^{i}$ is the observed correlation function (bootstrap) 
uncertainty.

In Fig. 5 we present, for various cosmological models, the variation of 
$\Delta \chi^{2}=\chi^{2}(b_{\circ})-\chi^{2}_{\rm min}(b_{\circ})$ 
around the best $b_{\circ}$ fit, for the different
richness class (left panel for $\rm S_{2}$
and right panel for $\rm S_{1}$).
\begin{figure}
\mbox{\epsfxsize=8cm \epsffile{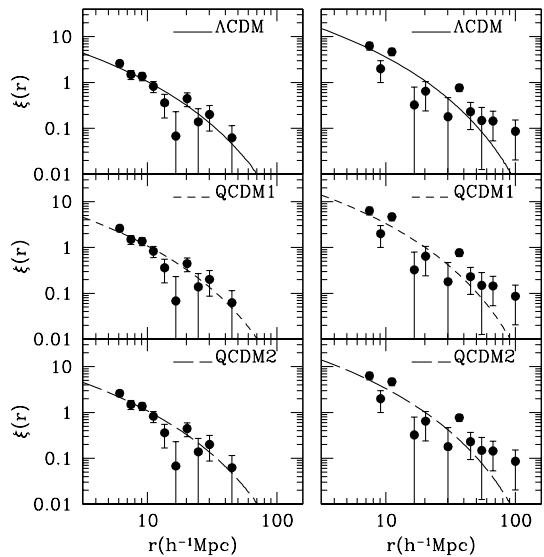}}
\caption{Comparison of the observed and model SDSS cluster correlation
  function: Left panel, $\rm S_{2}$ sample;
right panel, $\rm S_{1}$ sample. The observational 
data are represented by filled symbols.} 
\end{figure}

To this end, owing to the fact that the observational data are analyzed 
in redshift space, the correlations should be amplified by the
factor $K(\beta)=1+2\beta/3+\beta^{2}/5$ (Hamilton 1992) 
where $\beta\simeq \Omega_{\rm m}^{\cal{\alpha}}/b_{\circ}$.
We utilize the generic expression for ${\cal \alpha}$, 
defined by the Wang \& Steinhardt (1998):
\be\label{eq:26}
{\cal \alpha} \simeq \frac{3}{5-w/(1-w)}+
\frac{3}{125}\frac{(1-w)(1-3w/2)}{(1-6w/5)^{3}}(1-\Omega_{\rm m}) \;\;.  
\ee 
In Table 2 we list the results of the fits for our 
two cluster catalogs, ie., the
cosmological models and the value of the cluster optical bias, 
$b_{\circ}$, at the present time, as well as the redshift distortion 
$\beta$ parameter and a measure of the $K(\beta)$ correction. 
We find that the redshift space distortions effect 
increases $\xi_{\rm S_{j}}(r)$ by a factor of $\sim 12 - 26 \%$.

In Fig. 6, we plot the measured $\xi_{\rm S_{j}}(r)$ (filled symbols)
of our two samples 
with the estimated two point correlation
function for all three cosmological models.
We should conclude that the behavior of the observed two point  
correlation function of the galaxy clusters 
is sensitive to the different cosmologies with
a strong dependence on the present time bias.
By separating between low and 
high richness regimes, we obtain results being consistent with the 
hierarchical clustering scenario, in which the 
rich clusters are more biased tracers of the
underlying matter distribution with respect to the low richness
clusters. 

We can put some further cosmological constraints, comparing our 
clustering results with those based on large-scale dynamics.
For example Branchini \& Plionis (1996) using the cluster dipole 
after reconstructing the spatial distribution of Abell/ACO $R\ge 0$
clusters found $\beta_{\rm Abell}=0.21\pm 0.03$.
Also, Branchini at al. (2000) comparing the density and 
velocity fields of the Abell/ACO cluster distribution
with the corresponding POTENT fields (using
the MARK III galaxy velocity sample), obtained 
$\beta_{\rm POTENT}=0.22\pm 0.08$.
Comparing the latter $\beta$-results with our clustering predictions
(Table 2)
we can conclude that for the $\rm S_{1}$ sample (Abell $R\ge 0$ richness)
the only model which fails (although marginally) to reproduce the
large-scale dynamical results is QCDM2 ($w=-1/3$).

\section{Conclusions}
We have studied the clustering properties of the SDSS galaxy clusters 
in redshift space. We have divided the total sample in two richness
subsamples; 
roughly corresponding to Abell $R\ge 0$ ($N_{\rm gal}\ge 30$ members) 
and to APM ($N_{\rm gal} \ge 20$ members) clusters. We find 
that if the two point cluster correlation function
is modeled as a power law, $\xi(r)=(r_{\circ}/r)^{\gamma}$, then 
the best-fitting parameters are 
(a) $r_{\circ}=20.7^{+4.0}_{-3.8} \;h^{-1}$ Mpc with 
$\gamma=1.6^{+0.4}_{-0.4}$ and (b) $r_{\circ}=9.7^{+1.2}_{-1.2}
\;h^{-1}$ Mpc with $\gamma=2.0^{+0.7}_{-0.5}$ respectively.
We have also found that the Abell-like sample is not significantly 
affected by projection effects, and its apparent clustering elongation
along the line-of-sight is due to the survey geometry. However, the
APM-like sample appears to be somewhat affected by projection effects,
showing a clustering elongation along the line-of-sight larger than what
expected from the survey geometry.

Comparing the
cluster correlation function with the predictions of 3 
spatially flat quintessence models (having $\Omega_{\rm m}=0.3$), we estimate
the cluster redshift space distortion 
parameter $K(\beta)$ and we conclude
that the amplitude of the cluster redshift correlation function 
increases by a factor of $\sim 12-26 \%$ (depending on the richness class).
Finally, comparing our clustering results with those of dynamical
analysis, based on the large scale motions,
we find that the flat cosmological models with $w\le -0.6$ 
are consistent with the observational results.

\section* {Acknowledgments}
The $\Lambda$CDM simulation used in this paper was carried out by the Virgo 
Supercomputing Consortium using computers based at the Computing Centre of the
Max-Planck Society in Garching and at the Edinburgh parallel 
Computing Centre. The data are publicly
available at http://www.mpa-garching.mpg.de/NumCos.

This work is jointly funded by the European Union
and the Greek Government in the framework of the program 'Promotion
of Excellence in Technological Development and Research', project
{\em 'X-ray Astrophysics with ESA's mission XMM'}. Furthermore, MP
acknowledges support by the Mexican Government grant No
CONACyT-2002-C01-39679. Finally, we would like to thank the 
referee, Chris Collins, for useful suggestions.


{\small
\begin{thebibliography}{}
\bibitem[]{}Abadi, M. G., Lambas, D. G., Muriel, H., 1998, ApJ, 507, 526
\bibitem[]{}Bardeen, J.M., Bond, J.R., 
Kaiser, N. \& Szalay, A.S., 1986, ApJ, 304, 15
\bibitem[]{}Bagla J. S. 1998, MNRAS, 417, 424
\bibitem[]{}Bahcall, N., \&, Soneira, R. M., 1983, ApJ, 270, 20
\bibitem[]{}Bahcall, N., \&, Burgett, W. S., 1986, ApJ, 300, L35
\bibitem[]{}Bahcall, N., \&, West, M. J., 1992, ApJ, 392, 419
\bibitem[]{} Basilakos, S. \& Plionis, M., 2001, ApJ, 550, 522
\bibitem[]{} Basilakos, S. \& Plionis, M., 2003, ApJ, 593, L61
\bibitem[]{}Benson A. J., Cole S., 
Frenk S. C., Baugh M. C., \& Lacey G. C., 2000, MNRAS, 311, 793
\bibitem[]{}Borgani, S., Plionis, M., Kolokotronis, V., 1999, MNRAS, 305, 866
\bibitem[]{}Branchini, E., \&, Plionis. M., 1996, ApJ, 460, 569 
\bibitem[]{}Branchini, E., Zehavi, I., Plionis. M., 
\& Dekel, A., 2000, MNRAS, 313, 491
\bibitem[]{}Colberg, J.M., et al., 2000, MNRAS, 319 209
\bibitem[]{}Collins, C. A., et al., 2000, MNRAS, 319, 939
\bibitem[]{}Croft, R., A., C., Dalton, B., Efstathiou, G.,
 Sutherland, W., J., Maddox, S., J., 1997, MNRAS, 291, 305
\bibitem[]{}Dalton, B. G., Croft, R. A. C., Efstathiou, G., 
Sutherland, W. J., Maddox, S. J., Davis, M., 1994, MNRAS, 271, L47
\bibitem[]{}Efstathiou, G., Dalton, B. G.,  Sutherland, W. J., 
Maddox, S. J., 1992, MNRAS, 257, 125
\bibitem[]{}Freedman, W., L., et al., 2001, ApJ, 553, 47
\bibitem[]{} Frenk, C.S., et al, 2000, {\em astro-ph:0007362}
\bibitem[]{}Fry J.N., 1996, ApJ, 461, 65
\bibitem[]{} Fukugita, M., Shimasaku, K., Ichikawa, T. 1995, PASP,
107, 945
\bibitem[]{}Gonzalez, A. H., Zaritsky, D., Wechsler, R. H., 2002, ApJ, 571, 129
\bibitem[]{}Goto, T., et al., 2002, AJ, 123, 1825
\bibitem[]{}Hamilton, A. J. S., 1992, ApJ, 385, L5
\bibitem[]{}Hamilton, A. J. S., 1993, ApJ, 417, 19
\bibitem[]{} Jing, Y.P., Plionis, M., Valdarnini, R., 1992, ApJ, 389, 499
\bibitem[]{}Kaiser N., 1984, ApJ, 284, L9
\bibitem[]{}Kirkman, D., Tytler, D., Suzuki, N., O'Meara, J.M., Lubin,
D., 2003, ApJS submitted, {\em astro-ph/0302006} 
\bibitem[]{} Klypin, A. A. \& Kopylov, A. I., 1983, Sov.Astr.Let., 9, 41
\bibitem[]{}Lahav O., Edge, A. C., Fabian, A. C., 
Putney, A., 1989, MNRAS, 238, 881
\bibitem[]{}Mo, H. J., Jing, Y. P., B$\ddot {\rm o}$rner, G., 1992, 
ApJ, 392, 452   
\bibitem[]{}Moscardini, L., Matarrese, S., Mo, H. J., 2001, MNRAS, 327, 422
\bibitem[]{}Nichol, R. C., Briel, O. G., Henry, P. J., 1994, ApJ, 267, 771
\bibitem[]{}Nusser \& Davis, 1994, ApJ, 421, L1
\bibitem[]{}Olive, K.A., Steigman, G., Walker, T.P., Phys.Rep., 333, 389
\bibitem[]{}Peacock, A. J., \&, West, M., 1992, MNRAS, 259, 494
\bibitem[]{}Peacock, A. J., \&, Dodds, S. J., 1994, MNRAS, 267, 1020
\bibitem[]{}Peebles P.J.E., 1993. Principles of Physical Cosmology, 
Princeton University Press, Princeton New Jersey
\bibitem[]{}Peebles P.J.E., Ratra, B., 2003, {\em astro-ph/0207347}
\bibitem[]{}Plionis, M., \&, Basilakos, S., 2002, MNRAS, 329, L47
\bibitem[]{} Plionis, M., in {\em Cosmological Crossroads}, Springer
  LNP, 592, p.147 (eds. Cotsakis \& Papandonopoulos) 
\bibitem[]{}Sugiyama, N., 1995, ApJS, 100, 281
\bibitem[]{}Tago, E., Saar, E., Einasto, J., Einasto, M., 
Müller, V., Andernach, H., AJ, 123, 37 
\bibitem[]{}Tegmark M. \& Peebles P.J.E, 1998, ApJL, 500, L79
\bibitem[]{}Wang, L. \& Steinhardt, P.J., 1998, ApJ, 508, 483
\end{thebibliography}
}
\end{document}